\documentclass[%
 preprint,
 amsmath,amssymb,
 aps,
prl,
]{revtex4-2}

\usepackage{graphicx}% Include figure files
\usepackage{subcaption}
\usepackage{pgfplots}
\pgfplotsset{compat=1.18}
\usepackage{pgfplotstable}
\usepackage{dcolumn}% Align table columns on decimal point
\usepackage{bm}% bold math
\usepackage[hidelinks]{hyperref}
\usepackage{color} %to color the text
\usepackage{siunitx}
\DeclareSIUnit\eangstrom{\text{e\AA}}
\usepackage{amsmath}
\usepackage{braket}
\usepackage{mathtools}
\usepackage[version=4]{mhchem} %for chemical formulas

\usepackage[normalem]{ulem}
\usepackage[shortlabels]{enumitem}
\usepackage{enumerate}
\usepackage{enumitem}
\usepackage{fixltx2e} %to use textsubscript in section titles
\usepackage{upgreek}
\usepackage{booktabs}

\begin{document}

\section{Title}

Twist-angle Control of Nonlinear Interference in a ZnO Nanowire/Monolayer WSe2 Hybrid Structure

\section{Author list}
Maximilian Tomoscheit$^{1}$, Benedikt Mathes$^{1}$, Alexander Zaunick$^{1}$, Moritz Willems$^{1}$, Edwin Eobaldt$^{1}$, Priyanka S. Prakash$^2$, Eva Perlt$^{2}$, Carsten Ronning$^{1,3}$, and {Giancarlo} Soavi$^{1,3,*}$

\section{Affiliations}
\noindent
$^1$Institute of Solid State Physics, Friedrich Schiller University Jena, Helmholtzweg 5, 07743 Jena, Germany
\newline
$^2$Otto Schott Institute of Materials Research, Friedrich Schiller University Jena, Löbdergraben 32, 07743 Jena, Germany
\newline
$^3$Abbe Center of Photonics, Friedrich Schiller University Jena, Albert-Einstein-Straße 6, 07745 Jena, Germany
\newline
*Corresponding author: Giancarlo Soavi\\
email: giancarlo.soavi@uni-jena.de

\maketitle

\section{Abstract}
Nanoscale devices that integrate materials of different dimensionalities (0D, 1D, and 2D) hold great potential for advanced applications in photonics and optoelectronics. A fundamental requirement for the development of such devices is the engineering and control of light–matter interactions beyond the simple enhancement or quenching of linear and nonlinear optical emission. In this study, we demonstrate control over nanoscale light–matter interactions by achieving twist-angle tunability of the nonlinear optical response in a hybrid system composed of a ZnO nanowire and a monolayer of WSe$_2$. By varying the relative orientation between the ZnO polar axis and the WSe$_2$ crystal axes, we realize both constructive and destructive interference in second-harmonic generation, as well as full material selectivity in second harmonic polarization-dependent measurements. These outcomes arise from the distinct dimensionalities and symmetries of the hybrid constituents, underscoring the generality of our approach. Thus, our work presents an advanced framework for the design and control of nonlinear light-matter interactions in nanoscale hybrid devices, thereby paving the way for their future use in photonic and optoelectronic technologies.

\section{Introduction}
Nanoscale materials offer a versatile and powerful platform for applications in photonics and optoelectronics, as they hold the potential to substitute electronic devices, where information is transferred solely as electron currents, by photonic integrated circuits (PICs), where electrical signals are combined with light as a platform for logic operations~\cite{Zhang2025}. 

Within the realm of nanomaterials for PICs, two-dimensional layered materials, such as graphene and transition metal dichalcogenides (TMDs), play a leading role thanks to their ease of on-chip integration, and strong electrically tunable light-matter interactions~\cite{Roadmap2025, Photonics2025}. Several optoelectronic applications and devices based on layered materials have already been demonstrated, including ultrafast logical gates~\cite{Li2022, Zhang2022}, electrically tunable frequency converters~\cite{Soavi2018,Ghaebi2024}, and ultrafast valleytronic/spintronic devices~\cite{Herrmann2023, Friedrich2025, Herrmann2025b, Seyler2026}. 
Another key milestone for PIC is the engineering of nanoscale coherent light sources, a field where semiconducting nanowires (NWs) have emerged as ideal candidates~\cite{Eaton2016}. Owing to their geometry and high refractive index, NW lasers naturally provide waveguiding and light-amplification via total internal reflection, leading to lasing emission from their end facets~\cite{Ronning2010, Ronning2008}.
The hybridization of 2D layered materials with NW lasers can offer an exciting playground for fundamental research and PIC applications~\cite{Yan2024}. For instance, electrical tunability can be exploited to modulate the charge-transfer efficiency in hybrid TMD/NW lasers~\cite{Eobaldt2022}, thus achieving electrical tuning of the cavity losses and, as a consequence, of the lasing properties. This could lead to the photonic lasing analogs of what has so far been achieved only with plasmonic NW lasers coupled to graphene~\cite{Li2020, Li2019}.

Another interesting and promising research and technological direction for such NW-layered material hybrids is the modulation and engineering of their coupled optical properties, both in linear and nonlinear regimes of light-matter interactions.
For instance, Li et al. investigated the influence of strain on the nonlinear optical (NLO) response of a MoS$_2$ monolayer by placing it on a TiO$_2$ NW~\cite{Li2019}, reaching a second harmonic (SH) intensity enhancement by more than 2 orders of magnitude. In another recent work, Kim et al. reported a 50 times PL enhancement on an unstrained MoS$_2$ through hybridization with a TiO$_2$ NW placed in a nanogap~\cite{Kim2020}.
However, to date all experiments on such hybrid systems have been limited to the study of the enhancement or quenching of their linear and nonlinear optical response. In this work, we advance beyond these approaches by exploring and demonstrating both twist-angle-dependent SH interference and material specific SH polarization selectivity in a hybrid sample consisting of a ZnO NW and a monolayer WSe$_2$. These results are intrinsically general and of broad validity, since they arise from the differences in symmetries and optical properties of the two materials, combined with their nanoscale nature which lifts phase-matching constraints. 

\section{Nonlinear interference in hybrid NW-TMD samples}

\begin{figure}
    \centering
    \includegraphics[width=.7\linewidth]{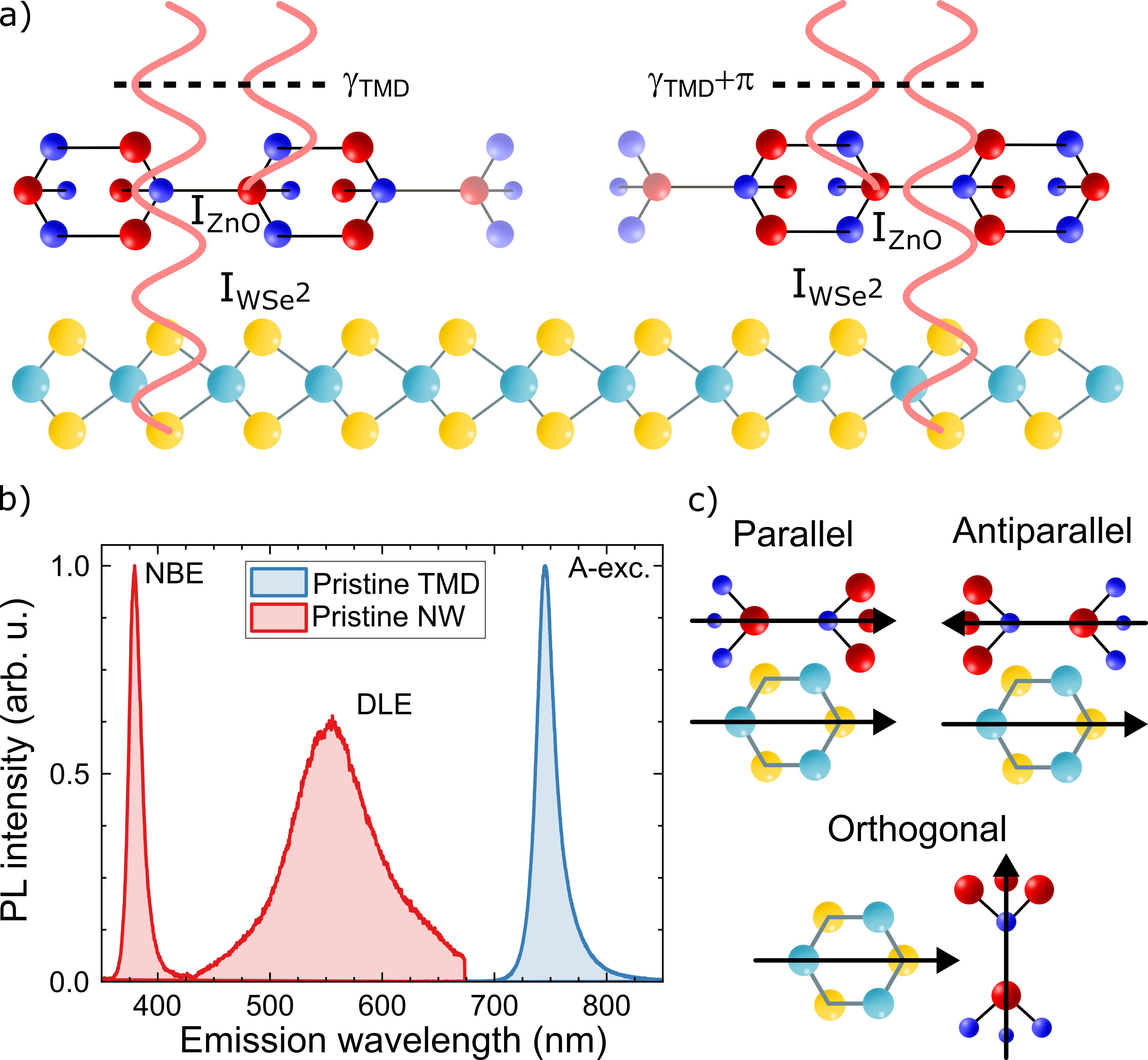}
    \caption{(a) Sketch of the SH interference in the NW-TMD hybrid structure for parallel (left) and antiparallel (right) dipole alignment.  (b) PL spectra of pristine ZnO NW (red) and TMD (blue). (c) Schematic illustration of the parallel, antiparallel and orthogonal dipole orientations, showing a single unit cell of each material with the dipoles indicated by a black arrow.}
    \label{fig:symmetries}
\end{figure}

The emitted SH intensity of any nonlinear hybrid structure will include interference terms in addition to the optical signal of the individual sample constituents. For a hybrid sample composed of two materials, the SH intensity is not simply proportional to the sum of the constituents second order susceptibilities $\chi_n^{(2)}$ but to the superposition
$I_\text{hyb}\propto|\chi^{(2)}_1 e^{i\gamma_1} +\chi^{(2)}_2 e^{i\gamma_2}|^2$, and thus contains an interference term defined by the phase difference $\Delta\gamma=\gamma_1-\gamma_2$. Such phase difference can originate from the twist-angle between the electric-dipoles of the different materials, from the complex nature of the NLO susceptibility $\chi^{(2)}$, which is purely imaginary at optical resonances and purely real in transparent regions, or from a combination of both~\cite{Paradisanos2022}. Our hybrid sample composed of a monolayer WSe$_2$ and a ZnO NW is an ideal platform to explore and exploit the combination of both sources of interference (see Fig.\ref{fig:symmetries}(a)). First, the optical resonances of WSe$_2$ and ZnO are spectrally well separated, as verified by photoluminesence (PL) spectroscopy performed on the pristine TMD and NW (see Fig.\ref{fig:symmetries}(b)). The ZnO NW spectrum shows two transitions, one near-band edge (NBE) emission at $\sim$ 380~nm and a much broader deep-level emission (DLE) centered at $\sim$ 550~nm, originating from deep defects\cite{Hou2014}. On the other hand, the TMD PL is dominated by the A-exciton emission at $\sim$ 750~nm. Thus, for resonant SHG on the lowest energy optical transition (A-exciton) of WSe$_2$, the ZnO NW will be fully transparent with a purely real $\chi^{(2)}$ NLO susceptibility. In this configuration, the SH interference directly probes the complex nature of the WSe$_2$ NLO susceptibility across the A-exciton resonance. 

Second, linear SH in TMDs (point group D$_\text{3h}$/$\bar{6}$m2) is emitted at an angle $-2\varphi$, where $\varphi$ is the relative angle between the fundamental beam (FB) polarization and the armchair (AC) crystal axis~\cite{Klimmer2021}. In TMDs there are three electric dipoles along the three equivalent AC directions pointing from the W to the Se atoms, leading to a negligible permanent dipole (see  Supporting Information S1 for DFT calculations of the electrical dipoles). In contrast, ZnO (point group C$_\text{6v}$/$\bar{6}$mm) is a polar material with a permanent dipole pointing from the O to the Zn atoms (see  Supporting Information S1)~\cite{Zvi2021}. The NW samples investigated in this study are grown such that the polar axis (c-axis, Wurtzite structure) coincides with the long (main) axis of the NW, while the nonpolar axis corresponds to its short axis. This leads to a strong anisotropy in the optical properties (including SHG) of the NW, with light emitted mainly/only along the NW main/polar axis. 

By combining and exploiting the NLO selection rules discussed above, we can thus engineer both SH interference and material selectivity by tuning the twist-angle between the TMD and the NW. In particular, we shall discuss three relevant configurations depicted in Fig.\ref{fig:symmetries}(c):
i) The NW polar axis is oriented along one of the three equivalent AC axes of the TMD, with electric dipoles having the same orientation (in the following this will be called the parallel configuration),
ii) In the antiparallel configuration, the NW polar axis is oriented along one of the TMD AC axes, with the corresponding electric-dipoles pointing in opposite directions. 
iii) Finally, we will discuss the orthogonal configuration, where the NW polar axis is oriented along one of the zigzag (ZZ) crystal axes of the TMD. 
In the following, we will show that the parallel and antiparallel configurations lead to both constructive and destructive SH interference, which we probe by tuning the FB wavelength across the A-exciton resonance of WSe$_2$. In contrast, the orthogonal configuration suppresses the interference term, while providing perfect material selectivity in polarization-dependent SH intensity maps.

For the experiments presented in this manuscript, we used mechanically exfoliated WSe$_2$ monolayers, transferred via PDMS onto a Si/SiO$_2$ substrate. To define the crystal orientation of the pristine TMD we performed polarization dependent SHG~\cite{Li2013}. The  Vapor-Liquid-Solid (VLS) grown ZnO NW was picked-up and positioned on top of the TMD monolayer at the desired angle (see Methods for details)~\cite{Eobaldt2022}. The NW was intentionally positioned with only half of its length on the TMD, to allow measurements on both pristine and hybrid parts of the sample. To realign the NW in different dipole orientations, the NW was picked up by PDMS, the sample was rotated and the NW was brought back into contact with the TMD.

\section{Twist-angle dependence of the NLO interference}
\begin{figure}
    \centering
    \includegraphics[width=.8\linewidth]{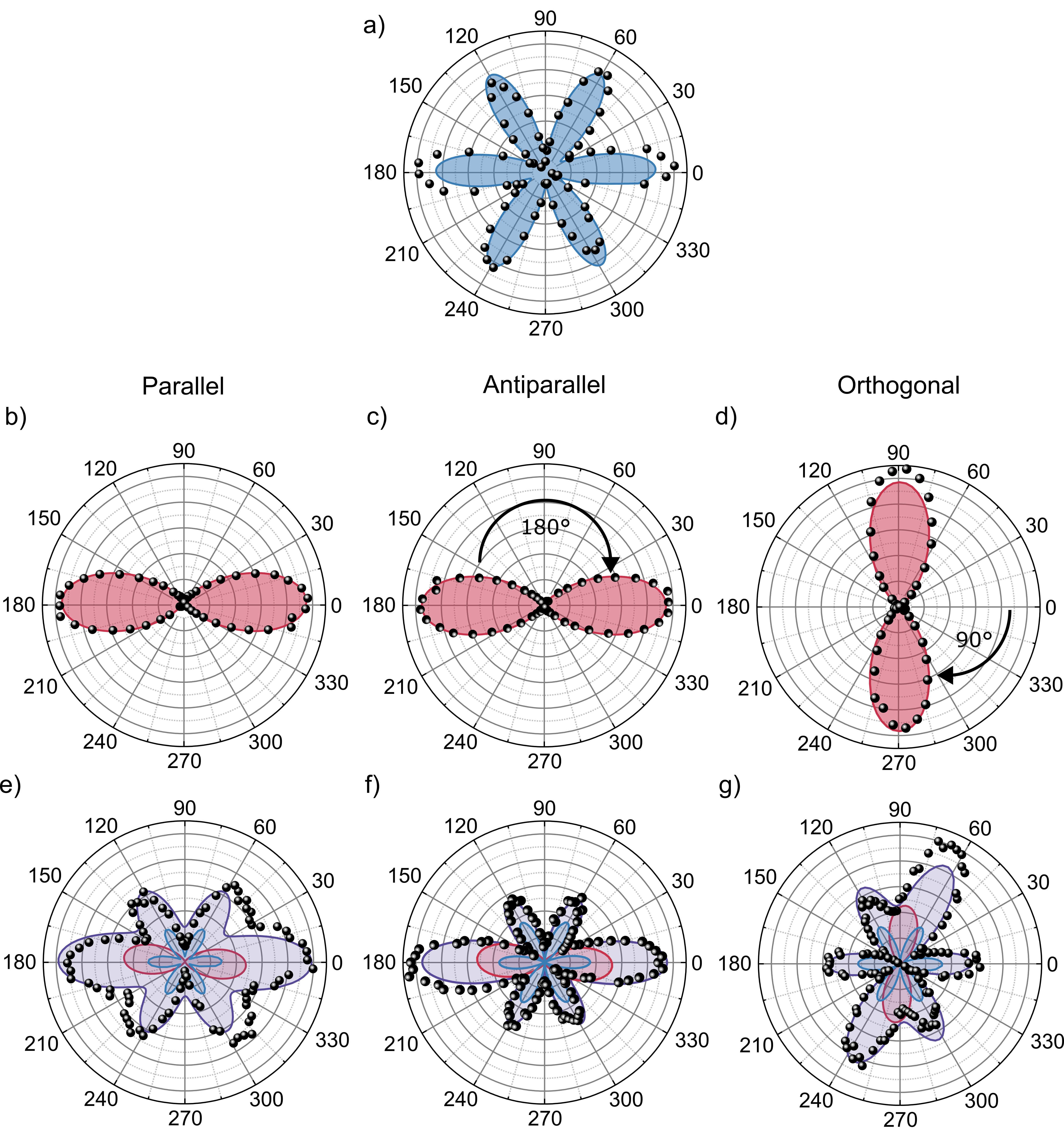}
    \caption{Polarization-dependent SHG in a co-polarized configuration measured for the TMD (panel a), NW (panels b-d), and hybrid structure (panels e-g). All polar plots are rotated with respect to the laboratory frame in order to have the TMD AC axis at zero angle. The shaded areas correspond to the fit performed using the equations described in the text.}
    \label{fig:polarization}
\end{figure}

We start our discussion from measurements of the SH polarization dependence of the hybrid NW-TMD sample and its pristine constituents in the parallel, antiparallel and orthogonal configurations (Fig.\ref{fig:symmetries}(c)). To do this, we performed co-polarized polarization-dependent SHG measurements. We filter and measure the SH intensity (via a polarizer/analyzer before the detector) that is emitted parallel to the polarization of the input fundamental beam field, which we control with a polarizer before the objective/sample (see Methods for details). All these measurements (Fig.\ref{fig:polarization}) were performed at a fixed FB wavelength of 1500~nm, and power of 20~mW. The data are normalized and rotated to have the TMD AC crystal axis at zero angle in the polar plots of Fig.\ref{fig:polarization}. 

For crystals belonging to the D$_\text{3h}$ point group, such as monolayer TMDs, there is only one non-vanishing element of the SH $\chi^{(2)}$ tensor, namely $\chi^{(2)}_\text{yyy}=-\chi^{(2)}_\text{xxy}=-\chi^{(2)}_\text{xyx}=-\chi^{(2)}_\text{yxx}=\chi^{(2)}_\text{TMD}$ \cite{Dogadov2022}, where x/y are the ZZ/AC axis of the TMD. This leads to the well-known polar-pattern ($I_\text{TMD}\propto\cos^2(3\varphi)$) for co-polarized polarization dependent measurements as shows in Fig.\ref{fig:polarization} (a), where the maxima correspond to the AC axis~\cite{Li2013}. 

For crystals belonging to the C$_\text{6v}$ point group, such as ZnO, there are four non-zero independent elements of the SH $\chi^{(2)}$ tensor, namely $\chi_\text{xzx}^{(2)}=\chi_\text{yzy}^{(2)}$, $\chi^{(2)}_\text{xxz}=\chi^{(2)}_\text{yyz}$, $\chi^{(2)}_\text{zxx}=\chi^{(2)}_\text{zyy}$ and $\chi^{(2)}_\text{zzz}$, where z is the polar axis \cite{Boyd2020}. However, in our experiments the diameter of the ZnO NW (150-250~nm) is much smaller than the FB wavelength (1500~nm), and thus all terms apart from $\chi^{(2)}_\text{zzz}$ are negligible. This leads to a strongly anisotropic SH intensity, where only co-polarized excitation/emission is allowed, namely $I_\text{NW}\propto |\chi^{(2)}_\text{NW}|^2 \cos^6(\phi)$, where $\phi$ is the relative angle between the FB polarization and the NW main/polar axis (Figs.\ref{fig:polarization}(b)-(d)). Here, we have defined $\chi^{(2)}_\text{zzz} = \chi^{(2)}_\text{NW}$ and, in addition, in our experiments the NW main/polar axis z lies in the xy plane of the TMD. We also notice that a rotation of 180 degrees of the NW has no effect on the SH intensity polarization dependence (see Fig.\ref{fig:polarization}(c)). In contrast, a rotation of the NW by 90° will rotate the SH emission accordingly (Fig.\ref{fig:polarization}(d)). 

For the hybrid system, the total SH intensity for any incident angle $I_\text{Hyb}(\psi)$ is clearly not simply the sum of the individual contributions, but it is further modulated due to interference. To model this, we write the expression of the total SH intensity for co-polarized measurements (see Supporting information S2 for details):

\begin{equation}
\begin{split}
I(\theta;\psi)\propto\left|\chi_\text{TMD}^{(2)}\right|^2\cos^2(3\theta+3\psi)+\left|\chi_\text{NW}^{(2)}\right|^2\cos^6\psi+2\chi_\text{TMD}^{(2)}\chi_\text{NW}^{(2)}\cos(3\theta+3\psi)\cos^3\psi\cos\gamma_\text{TMD}
\end{split}\label{eq:pol-gen}
\end{equation},
where $\psi$ is the relative angle between the FB polarization and the NW main/polar axis, $\theta$ is the relative angle between the TMD AC axis and the NW main/polar axis, and $\gamma_\text{TMD}$ is the complex phase of the nonlinear susceptibility of the TMD,  $\chi_\text{TMD}^{(2)} = |\chi_\text{TMD}^{(2)}|e^{i\gamma_\text{TMD}}$. We notice that for the parallel and antiparallel orientations ($\theta=0/\pi$) Eq.~\ref{eq:pol-gen} simplifies to

\begin{align}
\begin{split}
I(\theta=0/\pi;\psi)=I^{{\upharpoonleft\!\upharpoonright/{\downharpoonleft\!\upharpoonright}}}(\psi)\propto&\left|\chi_\text{TMD}^{(2)}\right|^2\cos^2(3\psi)+\left|\chi_\text{NW}^{(2)}\right|^2\cos^6(\psi)\\&\pm2\left|\chi_\text{TMD}^{(2)}\right|\left|\chi_\text{NW}^{(2)}\right|\cos(3\psi)\cos^3(\psi)\cos\gamma_\text{TMD}
\end{split}\label{eq:pol-par}
\end{align},
where the parallel/antiparallel orientations correspond to the $\pm$ sign of the interference term respectively. In contrast, the orthogonal orientation ($\theta=\pi/2$) leads to 

\begin{align}
\begin{split}
I(\pi/2;\psi)=I^\perp(\psi)\propto&\left|\chi_\text{TMD}^{(2)}\right|^2\sin^2(3\psi)+\left|\chi_\text{NW}^{(2)}\right|^2\cos^6(\psi)\\+&
2\left|\chi_\text{TMD}^{(2)}\right|\left|\chi_\text{NW}^{(2)}\right|\sin(3\psi)\cos^3(\psi)\cos\gamma_\text{TMD}
\end{split}\label{eq:pol-perp}
\end{align}.
The data in Fig.\ref{fig:polarization} panels (e)-(g) were fitted using Eq.~\ref{eq:pol-gen} to extract  the twist angle between the NW main/polar axis and the TMD AC axis. From the fits we retrieved twist angles of 4.3°$\pm$0.4° and 3.3°$\pm$0.5° for the parallel and antiparallel orientations respectively, and 85°$\pm$0.5° for the orthogonal orientation. Since these angles are very close to the ideal parallel (0), antiparallel ($\pi$) and orthogonal ($\pi /2$) orientations, we will use Eqs.~\ref{eq:pol-par} and~\ref{eq:pol-perp} in the following to demonstrate SH interference and material selective SH polarization maps. 

\section{Wavelength dependence of the NLO interference}

As discussed in the previous section, the SH interference can be further modulated by the complex nature and phase of the TMD NLO susceptibility. This is possible under the assumptions, which are verified under our experimental conditions, that: i) the FB and SH wavelengths are in a spectral region where the ZnO NW is transparent, leading to a purely real NLO susceptibility; ii) the SH wavelength is close to resonance for the TMD, leading to a complex NLO susceptibility; iii) co-polarized SHG measurements are performed for the parallel/antiparallel configurations by fixing the FB polarization along the AC axis of the TMD, corresponding to the main/polar axis of the NW. Under these hypothesis, namely $(\theta,\psi)=(0/\pi,0)$, we can rewrite Eq.~\ref{eq:pol-par} as

\begin{equation}
    I^{{\upharpoonleft\!\upharpoonright/{\downharpoonleft\!\upharpoonright}}}_\text{Hyb}(\lambda)\propto \left|\chi_\text{TMD}^{(2)}(\lambda)\right|^2+\left|\chi_\text{NW}^{(2)}(\lambda)\right|^2\pm2\left|\chi_\text{TMD}^{(2)}(\lambda)\right|\left|\chi_\text{NW}^{(2)}(\lambda)\right|\cos\left(\gamma_\text{TMD}(\lambda)\right)\label{eq:wavelength}
\end{equation}.
Eq.\ref{eq:wavelength} clearly shows that by fixing the relative angle between the NW main/polar axis and the TMD AC axis, and by fixing the FB polarization along the same direction, the SH interference is defined only by the complex phase $\gamma_\text{TMD}$ of the TMD NLO susceptibility close to the A-exciton resonance. 

\begin{figure}
    \centering
    \includegraphics[width=.9\linewidth]{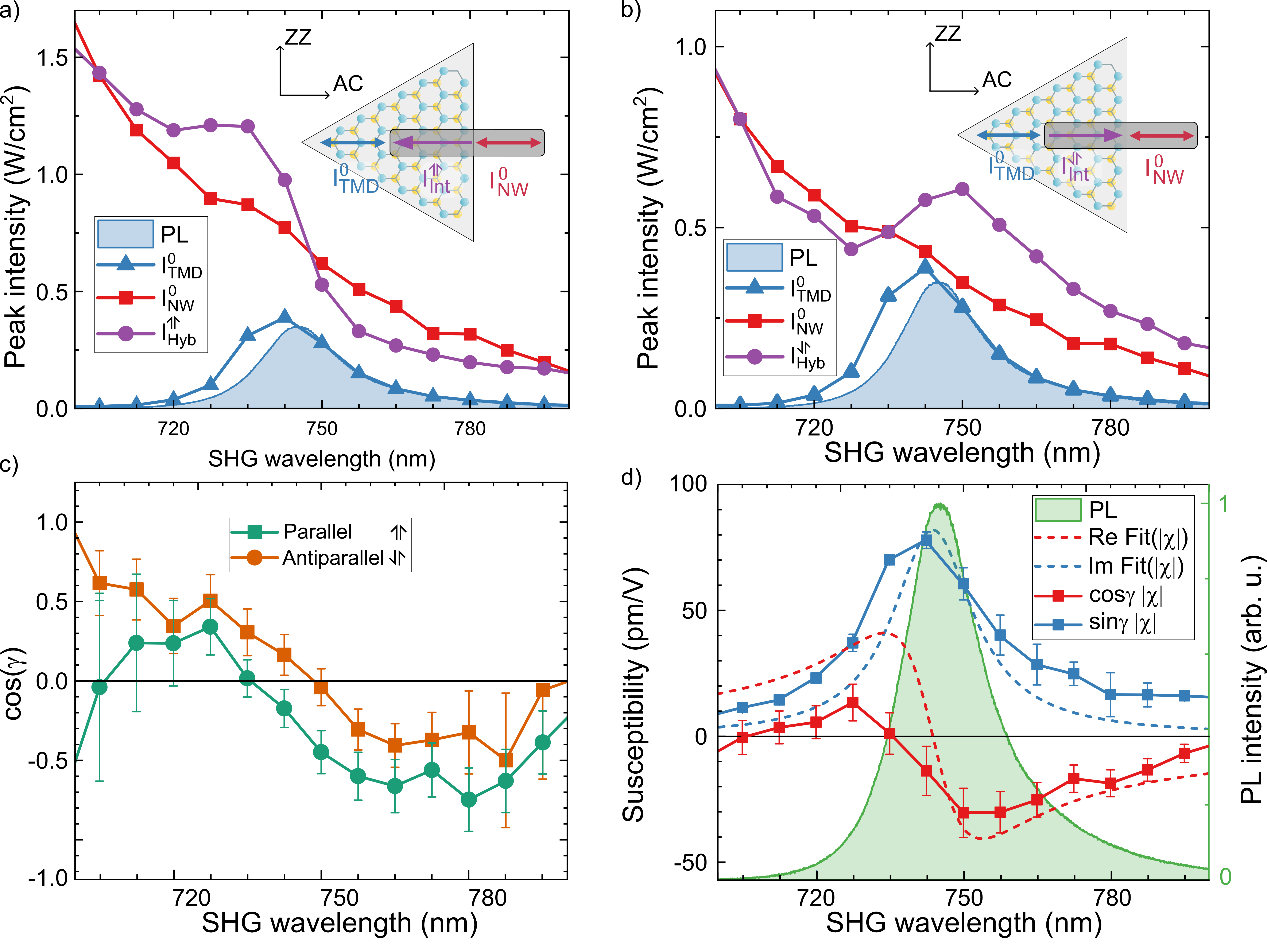}
    \caption{Wavelength-dependent SH intensity for the parallel (a) and antiparallel (b) orientations on the pristine TMD (blue triangles), pristine NW (red squares), and the hybrid structure (purple circles). The blue-shaded area is the WSe$_2$ PL spectrum measured for excitation with a 532~nm continues wave laser. (c) Complex phase of the TMD NLO susceptibility extracted from the equation and fit discussed in the main text. (d) Calculated TMD NLO susceptibility for the parallel configuration, following the procedure discussed in the main text. The PL spectrum of the TMD (green shaded area) is included for reference.}
    \label{fig:wavelength}
\end{figure}

To verify this, we performed experiments by comparing the PL (blue curve in Fig.\ref{fig:wavelength}(a,b)) and SH intensity as a function of the FB wavelength on the pristine TMD (blue triangles), pristine NW (red squares), and on the hybrid structure (purple circles). For these experiments, we tuned the SH wavelength across the A-exciton resonance of the TMD (FB wavelength from $\sim$ 1600~nm to $\sim$ 1400~nm). To account for inhomogeneities and thus SH intensity variations in the NW and hybrid structure, we rescaled the NW SH intensity to be identical to the SH intensity of the hybrid structure at a FB wavelength of 705~nm, where the SH intensity from the TMD is close to zero.

Fig.~\ref{fig:wavelength} (a) and (b) show the experimental results for the parallel and antiparallel configurations, respectively. The SH intensities of the pristine TMD and NW are almost identical in the two cases, if we account for small variations possibly due to inhomogeneities of the sample. The SH intensity of the TMD peaks (blue triangles) at approximately 742 nm, in correspondence to the PL emission of the A-exciton (blue curve). The slight blue shift of the SH peak compared to the PL emission could be due to bandgap modulation induced by the FB \cite{Klimmer2025}. At resonance, the measured SH intensity corresponds to an absolute value of the bulk susceptibility  $|\chi_\text{TMD}^{(2)}| \sim 80~\text{pm}/\text{V}$ (see Methods for details), in good agreement with the $90\pm10~\text{pm}/\text{V}$
values reported by Rosa et al. for WSe$_2$~\cite{Rosa2018}. The SH intensity of the NW (purple circles) decreases monotonically for increasing FB/SH wavelengths, as expected for transparent materials. 

However, the most interesting results are captured by the SH intensity of the hybrid sample (purple circles). As discussed, for both orientations the SH intensity is not simply the sum of the contributions from the two constituents (TMD and NW), but instead it displays a clear modulation in correspondence of the exciton resonance due to interference. In particular, the parallel orientation displays a local maximum at $\sim$ 727~nm and a local minimum at $\sim$ 765~nm, while the opposite trend is observed in the antiparallel orientation, namely a maximum at $\sim$ 750~nm and a minimum at $\sim$ 727~nm. This difference is due to the additional $\pi$ phase term introduced by rotating the NW main/polar axis.  
By re-arranging Eq.\ref{eq:wavelength} we can retrieve the complex phase term of the NLO susceptibility directly from our experimental results 
\begin{equation}
    \cos\gamma= \pm \frac{I_\text{Hyb}-I_\text{TMD}-I_\text{NW}}{2\sqrt{I_\text{TMD}I_\text{NW}}}\label{eq:phase}
\end{equation},
where the $\pm$ sign corresponds to the parallel/antiparallel configurations, respectively. The results are shown in Fig.\ref{fig:wavelength}(c). By combining the phase term and the absolute value, calculated from the SH intensity on the pristine TMD, we can reconstruct the complex NLO susceptibility of WSe$_2$ in the spectral region of the A-exciton resonance (see Methods for details), as shown in Fig.\ref{fig:wavelength}(d) with the blue (imaginary part) and red (real part) squares. In this graph, we also show the PL of pristine TMD (green), and the numerical results obtained by fitting the SH intensity of pristine TMD using a Lorentzian function and subsequently extracting its real and imaginary parts (dashed blue and red lines).

\section{Material Selective SHG by twist angle control and polarization filtering}

\begin{figure}
    \centering
    \includegraphics[width=\linewidth]{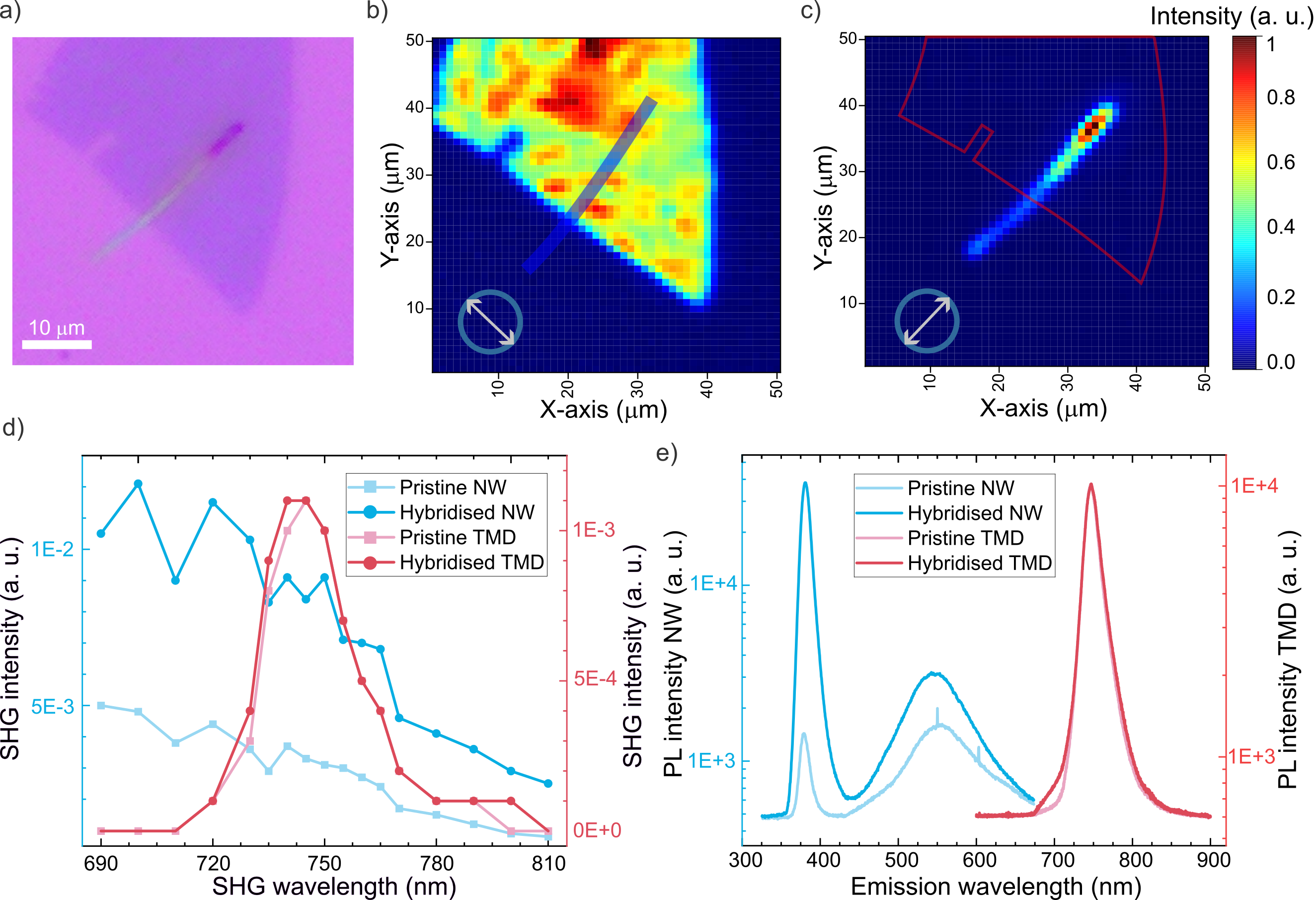}
    \caption{(a) Optical microscope image of the hybrid structure in the orthogonal orientation. (b, c) SHG maps for FB polarization parallel to the TMD AC axis (b) and NW main/polar axis (c). (d) SHG as a function of the SH wavelength for the pristine (light shades) and hybrid (dark shades) NW (blue) and monolayer WSe$_2$ (red). (e) Room-temperature PL spectra of pristine (light shaded) and hybrid (dark shades) NW (blue) and monolayer WSe$_2$ (red).}
    \label{fig:indep}
\end{figure}

Finally we discuss how to exploit the orthogonal alignment and co-polarized SH intensity measurements for material selective NLO mapping. As shown in Eq.\ref{eq:pol-perp}, we have two angles of incidence of the FB polarization in the orthogonal orientation for which the interference is fully suppressed: $(\theta,\psi)=(90°,0°)$ and $(\theta,\psi)=(90°,90°)$. In addition, Eq.\ref{eq:pol-perp} also shows that in the co-polarized configuration, excitation with the FB along the main/polar axis of the NW (corresponding to the ZZ axis of the TMD) leads to SH intensity emitted only by the NW. In contrast, when the FB polarization is along the TMD AC axis, the SH intensity from the NW is suppressed, and the signal only comes from the TMD.
These simple considerations can be used to achieve material selectivity in NLO experiments, as illustrated by the SH maps in Figs.~\ref{fig:indep} (b) and (c), while panel (a) shows a microscope image of the sample. The SH maps were acquired in a co-polarized configuration using a FB wavelength of 1500~nm, and by tuning the FB polarization from perpendicular (panel a) to parallel (panel b) with respect to the NW main/polar axis.

In addition to the material selectivity, the maps in Fig.\ref{fig:indep}(b) and (c) provide insights on the spatial distribution of the SH intensity in WSe$_2$ and the ZnO NW. The differences in the TMD SH intensity observed in panel (b) can be ascribed to spatial inhomogeneities of the sample, e.g. due to local strain, defects or doping \cite{Ghaebi2024,Carvalho2019,Mennel2018,Beach2020}, and they do not follow any specific trend. The maximum variation of SH intensity in this case is in the order of 50\%. In contrast, the differences in the SH intensity of the ZnO NW SHG follow a clear trend. First, the signal is enhanced at the end facets, possibly due to edge enhancement or waveguiding effects~\cite{Lai2022}. Second, the SH intensity from the portion of the NW lying on the TMD (hybrid structure) excluding the end facet is two times larger compared to the SH intensity from the pristine NW on the Si/SiO$_2$ substrate. The border between these two regions coincides with the edge of the TMD flake, as indicated by the red line in panel (c). 

To gain more insights into the observed spatial inhomogeneity of the NW SH intensity, we performed FB wavelength-dependent measurements also in the orthogonal configuration, as shown in Fig.\ref{fig:indep}(d). The enhancement is persistent over the entire FB wavelength range (from 1380~nm to 1620~nm) covered in our experiments, indicating that the effect does not arise from interference effects as in the case of the parallel and antiparallel configurations. In contrast, we do not observe any significant variation in the SH intensity emitted from the TMD in the pristine (light red) and hybrid (dark red) regions. These results are consistent with the PL intensity measured on the hybrid and pristine regions of the sample (Fig.\ref{fig:indep}(e)): also in this case we observe a reduction of the PL intensity from the NW in the hybrid compared to the pristine regions, while the PL from the TMD is almost identical in the two cases. While it is known that there is charge transfer in ZnO NW-TMD hybrid structures~\cite{Eobaldt2022}, a precise understanding of the physical origin of the changes in their PL and SH intensities is beyond the scope of this paper, and it will be subject of future investigations. However, the current work already clearly demonstrates that SHG is a powerful tool for non-invasive, material selective and ultrafast all-optical measurements on this family of hybrid samples based on constituents with mixed dimensionality.

\section{Conclusion}

We have demonstrated engineering, tunability and control of the nonlinear optical response of a hybrid nanoscale sample composed of a ZnO NW and a monolayer WSe$_2$. By aligning the NW main/polar axes along one of the AC crystal axis of the TMD, we were able to reconstruct the complex NLO susceptibility of WSe$_2$ in correspondence of the A-exciton resonance, and to modulate the SH intensity from the hybrid structure via interference effects. Alternatively, by aligning the NW main/polar axis along one of the ZZ axes of the TMD we were able to remove interference effects, while obtaining material selective SH maps. This configuration is also ideal to study and isolate the effects of hybridization (e.g., charge and/or energy transfer, bandgap renormalization etc.) on the individual constituents of the hybrid nanoscale device. On one hand, our results provide a viable route for future studies on the ultrafast interactions in NW-TMD hybrid structures. On the other hand, they offer new degrees of freedom for the engineering and control of nonlinear optics in technologically relevant nanoscale samples for integrated optics and photonics. 

\section{Methods}

\noindent
\textbf{Sample fabrication}
\\
The monolayer WSe$_2$ flake was mechanically exfoliated from a bulk synthetic crystal (HQ-graphene) and subsequently transferred to a Si/SiO$_2$ substrate \cite{Splendiani2010}.
The ZnO NW was synthesized using using the vapor-liquid-solid (VLS) mechanism based vapor-transport technique~\cite{Wagner1964}. The ZnO/graphite precursor, with a mass ratio of 7:1, was evaporated at a temperature of 1050°C within a three-zone tube furnace~\cite{Mueller2006}. The Ar/O$_2$ carrier gas, with a flow rate of 10 sccm, transports the vapor to a Si substrate with a 10-nm gold film, which is maintained at 1000 °C in the end zone of the furnace. The growth process was performed at 7 mbar for 1 hour~\cite{Eobaldt2022}.

For the manipulation of the NW position and twist angle a PDMS stripe, placed on a glass slide, was brought into contact with the NW. The PDMS was then quickly released from the substrate and the glass slide was positioned over the TMD flake on the target substrate. The target substrate was then rotated to achieve the desired relative orientation. After bringing the PDMS into contact with the TMD flake, the substrate was heated to 50°C to reduce the adhesion, and finally the glass slide was lifted up. The same procedure was repeated, for the same NW, to obtain the different twist angle configurations discussed in the main text (parallel, antiparallel and orthogonal).
\\
\noindent
\textbf{Error estimation}
\\
The error bars for the SH intensities in Fig.\ref{fig:wavelength}(c) and (d) were obtained considering two possible sources of errors: ($\Delta_\text{total}=\Delta_\text{sys}+\Delta_\text{exp}$): i) a systematic error,  extracted as the standard deviation from gaussian fits of the spectra and ii) experimental errors stemming from spatial inhomogeneity, which were extracted from the SHG maps in the orthogonal orientation (Fig.\ref{fig:indep}(b) and (c)). For the TMD, we extracted a section of the map, where the NW is placed, and calculated the standard deviation. For the NW, we extracted the maximum SH intensity from 7 line-scans perpendicular to the main axis. From these, we extracted the average and standard deviation for each FB wavelength. The errors for the phase and susceptibility (Fig.\ref{fig:wavelength}(c) and (d)) were calculated by error propagation.
\\
\noindent
\textbf{DFT calculations}
\\
All calculations employed Turbomole's~\cite{Turbomole} module riper~\cite{Continous-fast-Multipole-1} to perform periodic DFT calculations with explicit 2D and 3D periodic boundary conditions using local Gaussian basis functions.
The continuous fast multipole method \cite{Continous-fast-Multipole-1, Continous-fast-Multipole-2} with density fitting was applied.
For all calculations, the Perdew-Burke-Ernzerhof functional \cite{PBE} was applied in conjunction with the pob-TZVP-rev2 basis sets \cite{pob-TZVP-1, pob-TZVP-2}, which are designed for periodic calculations.
To account for London dispersion interactions, D3 dispersion correction \cite{D3-london-dispersion-correction} was applied in all calculations.
For the transition metal \ce{W}, 60 core electrons were represented by the corresponding effective core potentials to speed up calculations \cite{def-ecp}. Further details and parameters of the calculations as well as information on the structures are discussed in the Supporting Information.
\\
\noindent
\textbf{SH Intensity and NLO susceptibility}
\\
For each FB wavelength, we measure the number of counts per second at the detector. From this, we have to extract the number of emitted photons $N$ at the sample position by considering the losses of the setup (e.g., detector efficiency, transmission coefficients of filter and beamsplitters, etc.), which range from $\sim$ 90\% to $\sim$ 70\% depending on the wavelength. Finally, we can calculate the SH peak intensity as
\begin{equation}
    I_{2E}=\frac{2NeS}{\pi r_\text{Spot}^2ft_\text{pulse}}
\end{equation}
with $S=\sqrt{4\log2/\pi}$, the repetition rate $f=$ 67~MHz, the spot radius $r_\text{spot}= $3.11~$\upmu$m$/\sqrt{2}$ and the pulse duration $t_\text{pulse}= $200~fs$/\sqrt{2}$. The bulk SH tensor element $\left|\chi_\text{b}^{(2)}\right|$ can be calculated from the FB and SH peak intensities~\cite{Rosa2018}:
\begin{equation}
    \left|\chi_\text{bulk}^{(2)}\right|=\frac{\hbar}{dE}\cdot\sqrt{\frac{I_{2E}}{I_E}\frac{\varepsilon_0c^3}{32}(1+n_{2E})(1+n_E)^2}
\end{equation} 
where $I_E$ is the FB peak intensity, $I_{2E}$ the detected SH peak intensity, $n$ is the the refractive index at the FB and SH photon energies, and $d$~=~0.8 nm is the thickness of the WSe$_2$ monolayer~\cite{Li2014}.

\noindent
\textbf{SHG setup}
\\
For the FB laser source we used an optical parametric oscillator (OPO Levante IR from APE) pumped by an Yb-doped mode-locked laser (FLINT12, Light Conversion) with a repetition rate of 76~MHz and pulse duration of $\sim$ 100~fs. 
The FB was reflected by a Polka-Dot beam-splitter (BPD254-G, Thorloabs), and subsequently its polarization was tuned using a quarter-waveplate (SAQWP, B-Halle) combined with a linear polarizer (WP25M, Thorlabs). Afterwards, the FB was directed towards an objective (LMM-40X-P01, Thorlabs) and focused on the sample.  The back-reflected SH beam from the sample traveled the same optical path of the FB (linear polarizer and QWP) to obtain the desired co-polarized configuration discussed in the main text. Finally, the SH beam transmitted by the Polka-Dot beam-splitter was spectrally filtered and measured using either an avalanche photodiode (APD410A/M, Thorlabs) or a spectrometer (IHR320, Horiba).

\section{Acknowledgments}
The authors acknowledge funding from the Deutsche Forschungsgemeinschaft (DFG, German Research Foundation) via the SFB NOA 1375 A2 and B5 (project number 398816777), and WHAT-A-TWIST (project number 547611111).

\section{Data availability}
The data that supports the findings of this study are available from the corresponding author upon reasonable request.

\section{Ethics declarations}
The authors declare no competing interests.

\section{Author contribution}
G.S., C.R. and M.T. conceived the work. G.S. and M.T. developed the analytical model. M.T. and B.M. fabricated the hybrid samples, performed the measurements, and analyzed the data. E.E., M.W., A.Z., and C.R. fabricated and characterized the NWs. E.P and P.P. calculated the dipoles orientation. M.T. and G.S. wrote the manuscript with contributions from all authors.

\section{References}
\bibliographystyle{ieeetr}
\bibliography{interference}

@article{Turbomole,
    author = {Balasubramani, Sree Ganesh and Chen, Guo P. and Coriani, Sonia and Diedenhofen, Michael and Frank, Marius S. and Franzke, Yannick J. and Furche, Filipp and Grotjahn, Robin and Harding, Michael E. and Hättig, Christof and Hellweg, Arnim and Helmich-Paris, Benjamin and Holzer, Christof and Huniar, Uwe and Kaupp, Martin and Marefat Khah, Alireza and Karbalaei Khani, Sarah and Müller, Thomas and Mack, Fabian and Nguyen, Brian D. and Parker, Shane M. and Perlt, Eva and Rappoport, Dmitrij and Reiter, Kevin and Roy, Saswata and Rückert, Matthias and Schmitz, Gunnar and Sierka, Marek and Tapavicza, Enrico and Tew, David P. and van Wüllen, Christoph and Voora, Vamsee K. and Weigend, Florian and Wodyński, Artur and Yu, Jason M.},
    title = {TURBOMOLE: Modular program suite for ab initio quantum-chemical and condensed-matter simulations},
    journal = {J.~Chem.~Phys.},
    volume = {152},
    number = {18},
    pages = {184107},
    year = {2020},
    month = {05},
    issn = {0021-9606},
    doi = {10.1063/5.0004635},
    url = {https://doi.org/10.1063/5.0004635},
    eprint = {https://pubs.aip.org/aip/jcp/article-pdf/doi/10.1063/5.0004635/16685296/184107_1_online.pdf},
}

@article{Continous-fast-Multipole-1,
author = {Łazarski, Roman and Burow, Asbj{\"o}rn M. and Sierka, Marek},
title = {Density Functional Theory for Molecular and Periodic Systems Using Density Fitting and Continuous Fast Multipole Methods},
journal = {J.~Chem.~Theory~Comput.},
volume = {11},
number = {7},
pages = {3029-3041},
year = {2015},
doi = {10.1021/acs.jctc.5b00252},
URL = { https://doi.org/10.1021/acs.jctc.5b00252},
eprint = { https://doi.org/10.1021/acs.jctc.5b00252}
}

@article{Continous-fast-Multipole-2,
author = {Łazarski, Roman and Burow, Asbjörn Manfred and Grajciar, Lukáš and Sierka, Marek},
title = {Density functional theory for molecular and periodic systems using density fitting and continuous fast multipole method: Analytical gradients},
journal = {J.~Comput.~Chem.},
volume = {37},
number = {28},
pages = {2518-2526},
doi = {https://doi.org/10.1002/jcc.24477},
url = {https://onlinelibrary.wiley.com/doi/abs/10.1002/jcc.24477},
eprint = {https://onlinelibrary.wiley.com/doi/pdf/10.1002/jcc.24477},
year = {2016}
}

@article{pob-TZVP-1,
author = {Vilela Oliveira, Daniel and Laun, Joachim and Peintinger, Michael F. and Bredow, Thomas},
title = {BSSE-correction scheme for consistent gaussian basis sets of double- and triple-zeta valence with polarization quality for solid-state calculations},
journal = {J.~Comput.~Chem.},
volume = {40},
number = {27},
pages = {2364-2376},
doi = {https://doi.org/10.1002/jcc.26013},
url = {https://onlinelibrary.wiley.com/doi/abs/10.1002/jcc.26013},
eprint = {https://onlinelibrary.wiley.com/doi/pdf/10.1002/jcc.26013},
year = {2019}
}

@article{pob-TZVP-2,
author = {Peintinger, Michael F. and Oliveira, Daniel Vilela and Bredow, Thomas},
title = {Consistent Gaussian basis sets of triple-zeta valence with polarization quality for solid-state calculations},
journal = {J.~Comput.~Chem.},
volume = {34},
number = {6},
pages = {451-459},
doi = {https://doi.org/10.1002/jcc.23153},
url = {https://onlinelibrary.wiley.com/doi/abs/10.1002/jcc.23153},
eprint = {https://onlinelibrary.wiley.com/doi/pdf/10.1002/jcc.23153},
year = {2013}
}

@article{D3-london-dispersion-correction,
    author = {Grimme, Stefan and Antony, Jens and Ehrlich, Stephan and Krieg, Helge},
    title = {A consistent and accurate ab initio parametrization of density functional dispersion correction (DFT-D) for the 94 elements H-Pu},
    journal = {J.~Chem.~Phys.},
    volume = {132},
    number = {15},
    pages = {154104},
    year = {2010},
    month = {04},
    issn = {0021-9606},
    doi = {10.1063/1.3382344},
    url = {https://doi.org/10.1063/1.3382344},
    eprint = {https://pubs.aip.org/aip/jcp/article-pdf/doi/10.1063/1.3382344/15684000/154104_1_online.pdf},
}

@article{PBE,
  title = {Generalized Gradient Approximation Made Simple},
  author = {Perdew, John P. and Burke, Kieron and Ernzerhof, Matthias},
  journal = {Phys. Rev. Lett.},
  volume = {77},
  issue = {18},
  pages = {3865--3868},
  numpages = {0},
  year = {1996},
  month = {Oct},
  publisher = {American Physical Society},
  doi = {10.1103/PhysRevLett.77.3865},
  url = {https://link.aps.org/doi/10.1103/PhysRevLett.77.3865}
}

@article{def-ecp,
author = {Andrae, D. and Häußermann, U. and Dolg, M. and Stoll, H. and Preuß, H.},
year = {1990},
title = {Energy-adjusted ab initiopseudopotentials for the second and third row transition elements},
journal = {	Theor. Chim. Acta},
pages =  {123 - 141},
volume = {77},
issue = {2},
issn = {1432-2234},
url = {https://doi.org/10.1007/BF01114537},
doi = {10.1007/BF01114537},
}

@article{Li2014, title={Preparation and applications of mechanically exfoliated Single-Layer and Multilayer MOS2and WSE2Nanosheets}, volume={47}, url={https://doi.org/10.1021/ar4002312}, DOI={10.1021/ar4002312}, number={4}, journal={Accounts of Chemical Research}, author={Li, Hai and Wu, Jumiati and Yin, Zongyou and Zhang, Hua}, year={2014}, month=apr, pages={1067–1075} }

@article{Zhang2025,
    author = {Zhang, Yuning and Wu, Jiayang and Hu, Junkai and Jia, Linnan and Jin, Di and Jia, Baohua and Hu, Xiaoyong and Moss, David J. and Gong, Qihuang},
    title = {2D material integrated photonics: Toward industrial manufacturing and commercialization},
    journal = {APL Photonics},
    volume = {10},
    number = {4},
    pages = {040903},
    year = {2025},
    month = {04},
    issn = {2378-0967},
    doi = {10.1063/5.0249703},
    url = {https://doi.org/10.1063/5.0249703},
    eprint = {https://pubs.aip.org/aip/app/article-pdf/doi/10.1063/5.0249703/20491970/040903_1_5.0249703.pdf},
}

@article{Klimmer2025, title={Probing ultrafast coherent bandgap modulation in monolayer WSE 2 by nonlinear optics}, volume={14}, url={https://doi.org/10.1002/adom.202503236}, DOI={10.1002/adom.202503236}, number={5}, journal={Advanced Optical Materials}, author={Klimmer, Sebastian and Lettau, Thomas and Molina, Laura Valencia and Kartashov, Daniil and Peschel, Ulf and Wilhelm, Jan and Neshev, Dragomir and Soavi, Giancarlo}, year={2025}, month=dec }

@article{Carvalho2019, title={Nonlinear Dark-Field imaging of One-Dimensional defects in monolayer dichalcogenides}, volume={20}, url={https://doi.org/10.1021/acs.nanolett.9b03795}, DOI={10.1021/acs.nanolett.9b03795}, number={1}, journal={Nano Letters}, author={Carvalho, Bruno R. and Wang, Yuanxi and Fujisawa, Kazunori and Zhang, Tianyi and Kahn, Ethan and Bilgin, Ismail and Ajayan, Pulickel M. and De Paula, Ana M. and Pimenta, Marcos A. and Kar, Swastik and Crespi, Vincent H. and Terrones, Mauricio and Malard, Leandro M.}, year={2019}, month=dec, pages={284–291} }

@article{Splendiani2010, title={Emerging photoluminescence in monolayer MOS2}, volume={10}, url={https://doi.org/10.1021/nl903868w}, DOI={10.1021/nl903868w}, number={4}, journal={Nano Letters}, author={Splendiani, Andrea and Sun, Liang and Zhang, Yuanbo and Li, Tianshu and Kim, Jonghwan and Chim, Chi-Yung and Galli, Giulia and Wang, Feng}, year={2010}, month=mar, pages={1271–1275} }

@article{Mennel2018,
    author = {Mennel, Lukas and Paur, Matthias and Mueller, Thomas},
    title = {Second harmonic generation in strained transition metal dichalcogenide monolayers: MoS2, MoSe2, WS2, and WSe2},
    journal = {APL Photonics},
    volume = {4},
    number = {3},
    pages = {034404},
    year = {2018},
    month = {12},
    issn = {2378-0967},
    doi = {10.1063/1.5051965},
    url = {https://doi.org/10.1063/1.5051965},
    eprint = {https://pubs.aip.org/aip/app/article-pdf/doi/10.1063/1.5051965/14569661/034404_1_online.pdf},
}

@article{Beach2020, title={Strain dependence of second harmonic generation in transition metal dichalcogenide monolayers and the fine structure of the C exciton}, volume={101}, url={https://doi.org/10.1103/physrevb.101.155431}, DOI={10.1103/physrevb.101.155431}, number={15}, journal={Physical Review. B./Physical Review. B}, author={Beach, Kory and Lucking, Michael C. and Terrones, Humberto}, year={2020}, month=apr }

@article{Rosa2018, title={Characterization of the second- and third-harmonic optical susceptibilities of atomically thin tungsten diselenide}, volume={8}, url={https://www.nature.com/articles/s41598-018-28374-1}, DOI={10.1038/s41598-018-28374-1}, number={1}, journal={Scientific Reports}, author={Rosa, Henrique G. and Ho, Yi Wei and Verzhbitskiy, Ivan and Rodrigues, Manuel J. F. L. and Taniguchi, Takashi and Watanabe, Kenji and Eda, Goki and Pereira, Vitor M. and Gomes, José C. V.}, year={2018}, month=jun, pages={10035} }

@article{Herrmann2025b, title={Nonlinear valley selection rules and all-optical probe of broken time-reversal symmetry in monolayer WSe2}, volume={19}, url={https://www.nature.com/articles/s41566-024-01591-z}, DOI={10.1038/s41566-024-01591-z}, number={3}, journal={Nature Photonics}, author={Herrmann, Paul and Klimmer, Sebastian and Lettau, Thomas and Weickhardt, Till and Papavasileiou, Anastasios and Mosina, Kseniia and Sofer, Zdeněk and Paradisanos, Ioannis and Kartashov, Daniil and Wilhelm, Jan and Soavi, Giancarlo}, year={2025}, month=jan, pages={300–306} }

@book{Boyd2020, title={Nonlinear Optics}, journall={Academic}, author={R. W. Boyd}, year={2020}, publisher={Academic}}

@article{Mueller2006, title={Catalyst-Nanostructure interaction in the growth of 1-D ZNO nanostructures}, volume={110}, url={https://doi.org/10.1021/jp054476m}, DOI={10.1021/jp054476m}, number={4}, journal={The Journal of Physical Chemistry B}, author={Borchers, C. and M\"uller, S. and Stichtenoth, D. and Schwen, D. and Ronning, C.}, year={2006}, month=jan, pages={1656–1660} }

@article{Wagner1964,
    author = {Wagner, R. S. and Ellis, W. C.},
    title = {VAPOR‐LIQUID‐SOLID MECHANISM OF SINGLE CRYSTAL GROWTH},
    journal = {Applied Physics Letters},
    volume = {4},
    number = {5},
    pages = {89-90},
    year = {1964},
    month = {03},
    issn = {0003-6951},
    doi = {10.1063/1.1753975},
    url = {https://doi.org/10.1063/1.1753975},
    eprint = {https://pubs.aip.org/aip/apl/article-pdf/4/5/89/18416939/89_1_online.pdf},
}

@article{Li2019,
author = {Li, Dawei and Wei, Chengyiran and Song, Jingfeng and Huang, Xi and Wang, Fei and Liu, Kun and Xiong, Wei and Hong, Xia and Cui, Bai and Feng, Aixin and Jiang, Lan and Lu, Yongfeng},
title = {Anisotropic Enhancement of Second-Harmonic Generation in Monolayer and Bilayer MoS2 by Integrating with TiO2 Nanowires},
journal = {Nano Letters},
volume = {19},
number = {6},
pages = {4195-4204},
year = {2019},
doi = {10.1021/acs.nanolett.9b01933},
    note ={PMID: 31136188},

URL = { 
    
        https://doi.org/10.1021/acs.nanolett.9b01933
    
    

},
eprint = { 
    
        https://doi.org/10.1021/acs.nanolett.9b01933
    
    

}

}

@article{Kim2020,
author = {Kim, Jung Ho and Lee, Hyun Seok and An, Gwang Hwi and Lee, Jubok and Oh, Hye Min and Choi, Jihoon and Lee, Young Hee},
title = {Dielectric Nanowire Hybrids for Plasmon-Enhanced Light–Matter Interaction in 2D Semiconductors},
journal = {ACS Nano},
volume = {14},
number = {9},
pages = {11985-11994},
year = {2020},
doi = {10.1021/acsnano.0c05158},
    note ={PMID: 32840363},

URL = { 
    
        https://doi.org/10.1021/acsnano.0c05158
    
    

},
eprint = { 
    
        https://doi.org/10.1021/acsnano.0c05158
    
    

}

}

@Article{Klimmer2021,
author={Klimmer, Sebastian
and Ghaebi, Omid
and Gan, Ziyang
and George, Antony
and Turchanin, Andrey
and Cerullo, Giulio
and Soavi, Giancarlo},
title={All-optical polarization and amplitude modulation of second-harmonic generation in atomically thin semiconductors},
journal={Nature Photonics},
year={2021},
month={Nov},
day={01},
volume={15},
number={11},
pages={837-842},
issn={1749-4893},
doi={10.1038/s41566-021-00859-y},
url={https://doi.org/10.1038/s41566-021-00859-y}
}

@article{Li2013,
author = {Li, Yilei and Rao, Yi and Mak, Kin Fai and You, Yumeng and Wang, Shuyuan and Dean, Cory R. and Heinz, Tony F.},
title = {Probing Symmetry Properties of Few-Layer MoS2 and h-BN by Optical Second-Harmonic Generation},
journal = {Nano Letters},
volume = {13},
number = {7},
pages = {3329-3333},
year = {2013},
doi = {10.1021/nl401561r},
    note ={PMID: 23718906},

URL = { 
    
        https://doi.org/10.1021/nl401561r
    
    

},
eprint = { 
    
        https://doi.org/10.1021/nl401561r
    
    

}

}

@Article{Zvi2021,
author={Ben-Zvi, Regev
and Bar-Elli, Omri
and Oron, Dan
and Joselevich, Ernesto},
title={Polarity-dependent nonlinear optics of nanowires under electric field},
journal={Nature Communications},
year={2021},
month={Jun},
day={02},
volume={12},
number={1},
pages={3286},
issn={2041-1723},
doi={10.1038/s41467-021-23488-z},
url={https://doi.org/10.1038/s41467-021-23488-z}
}

@article{Dogadov2022, title={Parametric Nonlinear Optics with Layered Materials and Related Heterostructures}, volume={16}, url={https://doi.org/10.1002/lpor.202100726}, DOI={10.1002/lpor.202100726}, number={9}, journal={Laser \& Photonics Review}, author={Dogadov, Oleg and Trovatello, Chiara and Yao, Baicheng and Soavi, Giancarlo and Cerullo, Giulio}, year={2022}, month=jun }

@article{Roadmap2025,
author = {de Abajo, F. Javier García et al},
title = {Roadmap for Photonics with 2D Materials},
journal = {ACS Photonics},
volume = {12},
number = {8},
pages = {3961-4095},
year = {2025},
doi = {10.1021/acsphotonics.5c00353},

URL = { 
    
        https://doi.org/10.1021/acsphotonics.5c00353
    
    

},
eprint = { 
    
        https://doi.org/10.1021/acsphotonics.5c00353
    
    

}

}

@article{Herrmann2023, title={Nonlinear All‐Optical coherent generation and Read‐Out of valleys in atomically thin semiconductors}, volume={19}, url={https://doi.org/10.1002/smll.202301126}, DOI={10.1002/smll.202301126}, number={37}, journal={Small}, author={Herrmann, Paul and Klimmer, Sebastian and Lettau, Thomas and Monfared, Mohammad and Staude, Isabelle and Paradisanos, Ioannis and Peschel, Ulf and Soavi, Giancarlo}, year={2023}, month=may, pages={e2301126} }

@article{Paradisanos2022, title={Second harmonic generation control in twisted bilayers of transition metal dichalcogenides}, volume={105}, url={https://doi.org/10.1103/physrevb.105.115420}, DOI={10.1103/physrevb.105.115420}, number={11}, journal={Physical Review. B./Physical Review. B}, author={Paradisanos, Ioannis and Raven, Andres Manuel Saiz and Amand, Thierry and Robert, Cedric and Renucci, Pierre and Watanabe, Kenji and Taniguchi, Takashi and Gerber, Iann C. and Marie, Xavier and Urbaszek, Bernhard}, year={2022}, month=mar }

@article{Photonics2025, title={Photonics in Flatland: challenges and opportunities for nanophotonics with 2D semiconductors}, volume={2}, url={https://www.nature.com/articles/s44310-025-00092-3#article-info}, DOI={10.1038/s44310-025-00092-3}, number={1}, journal={Npj Nanophotonics}, author={Azimi, Ali et al}, year={2025}, month=dec }

@article{Li2022, title={Nonlinear co-generation of graphene plasmons for optoelectronic logic operations}, volume={13}, url={https://www.nature.com/articles/s41467-022-30901-8}, DOI={10.1038/s41467-022-30901-8}, number={1}, journal={Nature Communications}, author={Li, Yiwei and An, Ning and Lu, Zheyi and Wang, Yuchen and Chang, Bing and Tan, Teng and Guo, Xuhan and Xu, Xizhen and He, Jun and Xia, Handing and Wu, Zhaohui and Su, Yikai and Liu, Yuan and Rao, Yunjiang and Soavi, Giancarlo and Yao, Baicheng}, year={2022}, month=jun, pages={3138} }

@article{Zhang2022, title={Chirality logic gates}, volume={8}, url={https://doi.org/10.1126/sciadv.abq8246}, DOI={10.1126/sciadv.abq8246}, number={49}, journal={Science Advances}, author={Zhang, Yi and Wang, Yadong and Dai, Yunyun and Bai, Xueyin and Hu, Xuerong and Du, Luojun and Hu, Hai and Yang, Xiaoxia and Li, Diao and Dai, Qing and Hasan, Tawfique and Sun, Zhipei}, year={2022}, month=dec, pages={eabq8246} }

@article{Soavi2018, title={Broadband, electrically tunable third-harmonic generation in graphene}, volume={13}, url={https://www.nature.com/articles/s41565-018-0145-8}, DOI={10.1038/s41565-018-0145-8}, number={7}, journal={Nature Nanotechnology}, author={Soavi, Giancarlo and Wang, Gang and Rostami, Habib and Purdie, David G. and De Fazio, Domenico and Ma, Teng and Luo, Birong and Wang, Junjia and Ott, Anna K. and Yoon, Duhee and Bourelle, Sean A. and Muench, Jakob E. and Goykhman, Ilya and Conte, Stefano Dal and Celebrano, Michele and Tomadin, Andrea and Polini, Marco and Cerullo, Giulio and Ferrari, Andrea C.}, year={2018}, month=may, pages={583–588} }

@article{Ghaebi2024, title={Ultrafast Opto‐Electronic and thermal tuning of Third‐Harmonic generation in a graphene field effect transistor}, volume={11}, url={https://doi.org/10.1002/advs.202401840}, DOI={10.1002/advs.202401840}, number={31}, journal={Advanced Science}, author={Ghaebi, Omid and Klimmer, Sebastian and Tornow, Nele and Buijssen, Niels and Taniguchi, Takashi and Watanabe, Kenji and Tomadin, Andrea and Rostami, Habib and Soavi, Giancarlo}, year={2024}, month=jun, pages={e2401840} }

@article{Friedrich2025, title={Measurement of optically induced broken time-reversal symmetry in atomically thin crystals}, volume={20}, url={https://www.nature.com/articles/s41566-025-01801-2}, DOI={10.1038/s41566-025-01801-2}, number={2}, journal={Nature Photonics}, author={Friedrich, Florentine and Herrmann, Paul and Shanbhag, Shridhar Sanjay and Klimmer, Sebastian and Wilhelm, Jan and Soavi, Giancarlo}, year={2025}, month=nov, pages={186–193} }

@misc{Seyler2026, title={Valleytronics in 2D Materials Roadmap}, url={https://arxiv.org/abs/2603.01427}, journal={arXiv.org}, author={Seyler, Kyle L. and Soavi, Giancarlo and Weber, Bent and Das, Sunit and Agarwal, Amit and Paradisanos, Ioannis and Glazov, Mikhail M. and Dogadov, Oleg and Gucci, Francesco and Cerullo, Giulio and Conte, Stefano Dal and Biswas, Shubhadeep and Wilhelm, Jan and Žutić, Igor and Denisov, Konstantin S. and Zhou, Tong and Zheng, Huiyuan and Yao, Wang and Yu, Hongyi and Cao, Ting and Waters, Dacen and Yankowitz, Matthew and Burkard, Guido and Denisov, Artem and Ihn, Thomas and Ensslin, Klaus and Gaudreau, Louis and Boddison-Chouinard, Justin and Fedorova, Zlata and Staude, Isabelle and Goh, Kuan Eng Johnson and Zhou, Zhichao and Li, Xiao}, year={2026}, month=mar, language={en} }

@article{Eobaldt2022, title={Tuning nanowire lasers via hybridization with two-dimensional materials}, volume={14}, url={https://pubs.rsc.org/en/content/articlelanding/2022/nr/d1nr07931j}, DOI={10.1039/d1nr07931j}, number={18}, journal={Nanoscale}, author={Eobaldt, Edwin and Vitale, Francesco and Zapf, Maximilian and Lapteva, Margarita and Hamzayev, Tarlan and Gan, Ziyang and Najafidehaghani, Emad and Neumann, Christof and George, Antony and Turchanin, Andrey and Soavi, Giancarlo and Ronning, Carsten}, year={2022}, month=jan, pages={6822–6829} }

@article{Li2020, title={Current modulation of plasmonic nanolasers by breaking reciprocity on hybrid Graphene–Insulator–Metal platforms}, volume={7}, url={https://doi.org/10.1002/advs.202001823}, DOI={10.1002/advs.202001823}, number={24}, journal={Advanced Science}, author={Li, Heng and Huang, Zhen‐Ting and Hong, Kuo‐Bin and Hsu, Chu‐Yuan and Chen, Jia‐Wei and Cheng, Chang‐Wei and Chen, Kuo‐Ping and Lin, Tzy‐Rong and Gwo, Shang‐Jr and Lu, Tien‐Chang}, year={2020}, month=nov, pages={2001823} }

@article{Eaton2016, title={Semiconductor nanowire lasers}, volume={1}, url={https://www.nature.com/articles/natrevmats201628}, DOI={10.1038/natrevmats.2016.28}, number={6}, journal={Nature Reviews Materials}, author={Eaton, Samuel W. and Fu, Anthony and Wong, Andrew B. and Ning, Cun-Zheng and Yang, Peidong}, year={2016}, month=may }

@article{Yan2024, title={Semiconductor nanowire heterodimensional structures toward advanced optoelectronic devices}, volume={10}, url={https://pubs.rsc.org/en/content/articlelanding/2025/nh/d4nh00385c}, DOI={10.1039/d4nh00385c}, number={1}, journal={Nanoscale Horizons}, author={Yan, Xin and Li, Yao and Zhang, Xia}, year={2024}, month=oct, pages={56–77} }

@article{Ronning2010,
doi = {10.1088/0268-1242/25/2/024001},
url = {https://doi.org/10.1088/0268-1242/25/2/024001},
year = {2010},
month = {jan},
publisher = {},
volume = {25},
number = {2},
pages = {024001},
author = {Zimmler, Mariano A and Capasso, Federico and Müller, Sven and Ronning, Carsten},
title = {Optically pumped nanowire lasers: invited review},
journal = {Semiconductor Science and Technology}
}

@article{Ronning2008,
    author = {Zimmler, Mariano A. and Bao, Jiming and Capasso, Federico and Müller, Sven and Ronning, Carsten},
    title = {Laser action in nanowires: Observation of the transition from amplified spontaneous emission to laser oscillation},
    journal = {Applied Physics Letters},
    volume = {93},
    number = {5},
    pages = {051101},
    year = {2008},
    month = {08},
    issn = {0003-6951},
    doi = {10.1063/1.2965797},
    url = {https://doi.org/10.1063/1.2965797},
    eprint = {https://pubs.aip.org/aip/apl/article-pdf/doi/10.1063/1.2965797/14400388/051101_1_online.pdf},
}

@article{Hou2014,
    author = {Hou, Dongchao and Voss, Tobias and Ronning, Carsten and Menzel, Andreas and Zacharias, Margit},
    title = {Deep-level emission in ZnO nanowires and bulk crystals: Excitation-intensity dependence versus crystalline quality},
    journal = {Journal of Applied Physics},
    volume = {115},
    number = {23},
    pages = {233516},
    year = {2014},
    month = {06},
    issn = {0021-8979},
    doi = {10.1063/1.4884611},
    url = {https://doi.org/10.1063/1.4884611},
    eprint = {https://pubs.aip.org/aip/jap/article-pdf/doi/10.1063/1.4884611/15138615/233516_1_online.pdf},
}

@article{Lai2022,
    author = {Lai, Man-Hong and Chen, Wei-Liang and Lo, Chao-Yuan and Yu, Jia-Ru and Tang, Po-Wen and Chen, Chi and Chang, Yu-Ming},
    title = {The origin of edge-enhanced second harmonic generation in monolayer MoS2 flakes},
    journal = {AIP Advances},
    volume = {12},
    number = {10},
    pages = {105009},
    year = {2022},
    month = {10},
    issn = {2158-3226},
    doi = {10.1063/5.0104281},
    url = {https://doi.org/10.1063/5.0104281},
    eprint = {https://pubs.aip.org/aip/adv/article-pdf/doi/10.1063/5.0104281/16470044/105009_1_online.pdf},
}

\end{document}